\documentclass{ws-ijmpa}

\begin{document}

\markboth{Authors' Names}
{Instructions for Typing Manuscripts (Paper's Title)}

%
\catchline{}{}{}{}{}
%

\title{\bf THE DISCOVERY AND PROPERTIES OF PENTAQUARKS\\
}

\author{\footnotesize FL. STANCU
\footnote{
Plenary talk, MESON2004 Conference Proceedings, Crakow, June 4-8 2004
}}

\address{University of Liege, Physics Department,  \\
Sart Tilman, B-4000 Liege 1, Belgium
}

\maketitle


\begin{abstract}
The pentaquarks are exotic baryons formed of four quarks and an antiquarks.
Their existence has been discussed in the literature over the last 30 years 
or more, 
first in connection with kaon nucleon scattering data. The subject
has been revived by the end of 2002 when
experimental evidence of a narrow baryon of strangeness $S$ = + 1, and mass
$M \simeq$ 1530 MeV
has been found. This is  interpreted as the 
lightest member 
of an SU(3)-flavor antidecuplet. 
Here we shall mainly review the predictions of pentaquark 
properties as e.g. mass, spin and parity, within constituent quark models.
 Both light and heavy 
pentaquarks will be presented.

\keywords{Pentaquarks; parity and spin; constituent quark models.}
\end{abstract}

\section{Historical Note and the Present Experimental Observation}	
The possible existence of exotic hadrons forming a baryon antidecuplet 
with spin 1/2 and positive 
parity has been mentioned in the literature even before the advent of
QCD, in connection with the $KN$ scattering data. \cite{GOLOWICH}
Later on, the existence of multiquark systems appeared as natural in 
QCD. Several constituent quark model calculations were performed,
for example in Refs. \refcite{JAFFE,SORBA,STROTTMAN}. As a consequence,
searches were made in the 1.74-2.16GeV/$c^2$ mass range 
(for a review see Ref.~\refcite{OHASHI}) and 
the lowest state was thought to have negative parity.  At that time 
these exotic hadrons carried the name of Z* resonances. They were reviewed
by the Particle Data Group (PDG) until 1986, when they were suppressed
from the listings due to poor experimental evidence. However a new wave of 
theoretical
interest appeared soon after, in the context of the chiral soliton or the Skyrme
model. The first estimate of the lightest pentaquark mass, presently
named $\Theta^+$, was given by Praszalowicz \cite{PRASZ},
with a mass of the order 1.5 GeV .  Ten years later
Diakonov, Petrov and Polyakov \cite{DPP}, predicted not only a mass of a 
similar value but also a strong decay width  not larger
than 15 MeV, which means small on the hadronic scale. 
The pentaquark  $\Theta^+$ was identified as the lightest member of an
SU(3)-flavor antidecuplet having positive parity and spin 1/2. The 
predictions of Ref. \refcite{DPP} have motivated and oriented new experimental
searches of pentaquarks leading to the observation of a narrow resonance
at about the predicted mass \cite{NAKANO,BARMIN}. So far this observation
has been confirmed by another nine experimental groups \cite{EXP}, with various
projectiles and targets.
There are however a few experiments where $\Theta^+$ has not been
seen \cite{NEGEXP}.
Although $\Theta^+$ is now listed as a three-star resonance by PDG,
one is perfectly aware that individual observations have limited statistics
and further confirmation is desirable \cite{HICKS}.
  
The interpretation of  $\Theta^+$  as a pentaquark $uudd \bar s$ was
strengthened by the observation of another two narrow resonances
at an invariant mass of about 1862 MeV, in the $\Xi^- \pi^+$
and $\Xi^- \pi^-$ channels of the $p+p$ scattering \cite{NA49}.
They were candidates for the  $\Xi^0_{3/2}$ and   $\Xi^{--}_{3/2}$ 
members of the antidecuplet, having a quark content $udss \bar u$
and $ddss \bar u$ respectively. However, no evidence for the $\Xi(1862)$
resonances has been found by other three experiments
\cite{DIS04,WA89}. Thus the existence of  $\Xi(1862)$ remains entirely
controversial.

At the time when the observation of light pentaquarks became hopeless,
theoretical predictions were oriented towards the heavy sector. Based on
general arguments, the expectation was that that heavy pentaquarks would
be more stable than the light ones, thus easier to be observed. 
Simultaneously, two independent studies \cite{JMR} based on the one-gluon
exchange hyperfine interaction predicted stable 
charmed strange pentaquarks of content $uuds \bar c$ or $udds \bar c$.
The searches made for these pentaquarks at Fermilab remained however
inconclusive \cite{AITALA}. These were negative parity pentaquarks. 
By this time, in the context of a constituent quark model based on a 
pseudo scalar meson exchange hyperfine interaction \cite{GR,GPP}, 
heavy positive parity pentaquarks were proposed \cite{FS1}. 
In the charm sector the content of the
lowest pentaquark was $uudd \bar c$ with spin 1/2 or 3/2. Early this year 
the H1 Collaboration reported  a narrow resonance at 3099 MeV which
was interpreted as an anticharmed baryon with a
minimal content $uudd \bar c$ of spin 1/2 or 3/2 \cite{AKTAS}. 
However such a signal was not observed in a preliminary ZEUS analysis of 
$e-p$ collisions \cite{ZEUS}.
\section{\bf Present Approaches}

It would be difficult to make a review of 250  manuscripts or more,
posted on the LANL archives since July 1st 2003 (part of them 
already published). On the theoretical side one could merely
make an inexhaustible  list of the subjects under study. Some of these are:
\begin{itemlist}
\item
Determination of spin and parity of $\Theta^+$ 
(polarization experiments)
\item
Consistency between the calculated and/or observed limit on the width of
$\Theta^+$ and the $KN$ partial wave analysis
\item
Calculation of the photo-production cross sections on proton and neutron,
useful in determining the yet unknown production mechanism of $\Theta^+$ 
\item
The chiral soliton model revisited, limits on masses $\&$ widths of
the antidecuplet members
\item
The Skyrme model revisited (bound state or rigid rotator)
\item
Group theoretical classification of 
$ q \times q \times q \times q \times \bar q$ states and  
mass formulae
\item
The pentaquark $\Theta^+$ and its antidecuplet partners in constituent quark 
models
\item
The octet-antidecuplet or higher representation mixing
\item
Interpretation of $\Theta^+$ as a heptaquarks or as $K \pi N$ molecule
\item
The description of pentaquarks in the instanton model
\item
Pentaquark results from QCD sum rules
\item
Pentaquarks in lattice calculations
\item
Magnetic moments of pentaquarks
\item
$\Theta^+$ in relativistic heavy ion collisions
\end{itemlist}

The chiral soliton model describes $\Theta^+$ 
as a collective excitation of the mean chiral field in the spin and
isospin space. 
That is considered as the main reason for the low mass
and the very small  width of $\Theta^+$. In the chiral soliton model,
$\Theta^+$ and its partners form 
an antidecuplet with J$^P$ = 1/2$^+$, all being narrow resonances \cite{DPP}. 
The predictions of the chiral soliton model
for the masses and widths of the antidecuplet have recently been re-analyzed
\cite{EKP}. The mass ranges   are estimated to be 
1430 MeV $< M(\Theta^+) < $ 1660 MeV and 
1790 MeV $< M(\Xi^{--}) < $ 1970 MeV and the width of $\Theta^+$ remains
small.

The chiral soliton model is more fundamental, it naturally incorporates 
relativistic effects, but it is more difficult to apply  to hadron
spectroscopy. Contrary, the constituent quark model is essentially 
phenomenological, but it is more intuitive and more appropriate to describe 
spectra and decay of baryons and mesons. 
The two models are rather complementary. Then the question is
whether or not constituent quark models can accommodate the pentaquark
antidecuplet predicted by the chiral soliton model.

In a naive estimate, the nucleon mass is approximately the sum 
of masses of the constituent quarks. Taking  $m_u$ = $m_u$ = 315 MeV 
one gets 945 MeV for the nucleon. In a similar way 
the constituent quark model
gives  $M(\Theta^+)$  = 4 $m_u$ + $m_s \simeq $ 1700 MeV for a mass
difference $m_s - m_u \simeq $  150 MeV. This value of  $M(\Theta^+)$
is larger than the original chiral soliton model prediction and
the present average of the experimental value.

However in a proper  calculation
the total mass results from the above free mass term plus contributions from 
the kinetic energy, the confinement potential and the (short range) hyperfine
interaction.   
Then the main issues in any quark model are:
\begin{itemlist}
\item
The spin and parity of  $\Theta^+$ 
\item
The absolute mass of  $\Theta^+$ 
\item
The splitting between the isomultiplets of the antidecuplet
\item
The strong decay width
\item
The role of the SU(3)-flavor mixing representations.
\end{itemlist}

In the following we shall address these questions.  
There are two widely used constituent quark models:
the one-gluon exchange model where the hyperfine interaction has a
color-spin (CS) structure and the pseudoscalar meson 
(or Goldstone boson) exchange, where the hyperfine interaction has a
flavor-spin (FS) structure.  Below these models will be often
called the CS and FS models. 

The FS model gives a good description of
the baryon spectra, reproducing the correct order of positive and negative
parity levels in low energy spectra of both non-strange and strange baryons,
in contrast to the CS model.
But this model does not provide a description of the meson spectra
as the CS model does. The FS model gets some support from the
large $N_c$ QCD limit, where the exact symmetry is the flavor-spin
symmetry.
\section{\bf The Parity and Spin of the Pentaquark $\Theta^+$ in Constituent 
Quark Models}

A pentaquark state, described as a $q^4 \bar q$ system can be obtained 
from the direct product of a baryon ($q^3$) and a meson ($q \bar q$) state. 
In the flavor space this involves the direct product 
\begin{equation}\label{REPR}
8_F \times 8_F = 27_F + 10_F +{\overline {10}}_F + 2(8)_F + 1_F
\end{equation}
which shows that the antidecuplet  ${\overline {10}}_F$  is one of the possible
multiplets. 

An important issue is the parity and spin of a pentaquark antidecuplet. 
The parity is given by  $P = (-)^{\ell + 1}$  
where $\ell$ is the angular momentum of the system and - 1 the parity of 
the antiquark. Thus to obtain a positive parity the whole system 
must contain at least one (or an odd number) of units of angular excitation.

Few years ago,  based on group theory arguments,
it has been shown that the lowest state of heavy pentaquarks 
has positive parity \cite{FS1} in the FS model. The proof can 
be straightforwardly extended to light pentaquarks.
In a similar way one can show that the CS model also leads to
positive parity for the lowest state when the $q^4$ subsystem  has
isospin $I = 0$ \cite{MJ}, compatible with the content 
$uudd \bar s$ of $\Theta^+$ . In both models 
the hyperfine attraction is large enough to overcome the excess 
of kinetic energy brought by the excitation of a quark to the $p$-shell
and that is why the positive parity appears below the negative parity
state which does not contain any orbital excitation.

\section{\bf Dynamical Calculations for the Mass Spectrum in the Flavor-Spin
Model}
To our knowledge, 
there are practically no dynamical calculations for the pentaquark
antidecuplet in a Hamiltonian model containing a 
CS hyperfine interaction. The literature is restricted to some attempts 
based either on a schematic \cite{CHEUNG,HS} CS interaction
\begin{equation}\label{CS}
 V_{CS} = -  \sum_{i<j}^5 C^{CS}_{ij} \lambda_{i}^{c} \cdot \lambda_{j}^{c}
\vec{\sigma}_i \cdot \vec{\sigma}_j ,
\end{equation}
where all spatial variables have been integrated out 
and the parameters  $C^{CS}_{ij}$ are fitted to ordinary baryons,
or on simple models 
where the existence of correlated
diquark pairs in the orbital space is postulated, but not dynamically 
demonstrated \cite{DUDEK}. In the latter case
there is no antisymmetrization between quarks belonging to
different diquarks. In a related problem,
as for example the nucleon-nucleon interaction,
the antisymmetrization between quarks belonging to different nucleons 
has been proved crucial in describing the 
potential at short distances, see e. g. \cite{OKA,SPG}. There is no reason  
to neglect the antisymmetrization in $q^4$ subsystems.

Below we present dynamical calculations in the FS model. For the
$q^4$ subsystem we use the following Hamiltonian \cite{GPP}

\begin{equation}\label{ham}
H= \sum_i m_i + \sum_i \frac{\vec{p}_{i}^{\,2}}{2m_i} - 
\frac {\vec{P}^2}{2 M} + 
\sum_{i<j} V_{conf}(r_{ij}) + \sum_{i<j}
V_\chi(r_{ij})~,
\nonumber
\end{equation}

\begin{equation}\label{conf}
 V_{conf}(r_{ij}) = 
 -\frac{3}{8}\lambda_{i}^{c}\cdot\lambda_{j}^{c} \, C
\, r_{ij}~, 
\nonumber
\end{equation}

\begin{eqnarray}\label{VCHI}
V_\chi(r_{ij})
&=&
\left\{\sum_{F=1}^3 V_{\pi}(r_{ij}) \lambda_i^F \lambda_j^F \right.
\nonumber \\
&+& \left. \sum_{F=4}^7 V_{K}(r_{ij}) \lambda_i^F \lambda_j^F
+V_{\eta}(r_{ij}) \lambda_i^8 \lambda_j^8
+V_{\eta^{\prime}}(r_{ij}) \lambda_i^0 \lambda_j^0\right\}
\vec\sigma_i\cdot\vec\sigma_j~.
\nonumber 
\end{eqnarray}
\noindent
The analytic form of  $V_\gamma (r)$  
($\gamma = \pi, K, \eta$ or $\eta'$)  is 
   
\begin{equation}\label{RADIAL}
V_\gamma (r)=
\frac{g_\gamma^2}{4\pi}\frac{1}{12m_i m_j}
\{\theta(r-r_0)\mu_\gamma^2\frac{e^{-\mu_\gamma r}}{ r}- \frac {4}{\sqrt {\pi}}
\alpha^3 \exp(-\alpha^2(r-r_0)^2)\}~,
\end{equation}
with the parameters:
\begin{eqnarray}\label{PARAM}
&\frac{g_{\pi q}^2}{4\pi} = \frac{g_{\eta q}^2}{4\pi} =
\frac{g_{Kq}^2}{4\pi}= 0.67,\,\,
\frac{g_{\eta ' q}^2}{4\pi} = 1.206, &\nonumber\\ 
&r_0 = 0.43\,~ \mathrm{fm}, ~\alpha = 2.91 \,~ \mathrm{fm}^{-1}, 
C= 0.474 \, { fm}^{-2}, \, 
  m_{u,d} = 340 \,~ \mathrm{MeV}, \, m_s = 440 \,~ \mathrm{MeV}, \\
&\mu_{\pi} = 139 \,~ \mathrm{MeV},~ \mu_{\eta} = 547 \,~ \mathrm{MeV}.~
\mu_{\eta'} = 958 \, \mathrm{MeV},~ \mu_{K} = 495 \,~ \mathrm{MeV}.&
\nonumber
\end{eqnarray}
\noindent which lead to a good description of low-energy
non-strange and strange baryon spectra. Fixing the
nucleon mass at $m_N$ = 939 MeV, this parametrization gives
for example  $m_{\Delta}$ = 1232 MeV and the Roper resonance  
$N(1440)$ at 1493 MeV.  The lowest negative 
parity states  $N(1535)$ and $N(1520)$ appear at 1539 MeV, i. e.
above the Roper resonance, in agreement with the experiment.

Note that in this model the SU(3)$_F$ symmetry is broken  due to the
the mass  difference between the $s$ and the $u$ or $d$ quarks and
due to the differences in the pseudoscalar meson masses.

For the light pentaquark antidecuplet of which $\Theta^+$ is a member,
the above Hamiltonian must be supplemented by a term containing 
a $q \bar q$ interaction. In Ref. \refcite{SR}
this interaction was chosen to be spin dependent, but flavor independent.
Its schematic form was 
\begin{equation}\label{SPINSPIN}
{V}_{q \bar q }\ =\ \ {V}_{0} \ \sum\limits_{i}^{4}
{\vec{\sigma }}_{i} \cdot {\vec{\sigma}}_{\overline s}.
\end{equation}
Here $V_0$ is a phenomenological constant, which should correspond to
the ground state matrix element of the spin-spin part
of the $\eta$-meson exchange interaction. The role of the
interaction (\ref{SPINSPIN}) is to lower
the energy of the whole system towards stability.
In Ref. \refcite{FS2} it was assumed that an interaction
of type (\ref{SPINSPIN}) lowers all members of the antidecuplet
by the same amount which was fixed such as to reproduce the
mass of $\Theta^+$.


\begin{table}[h]
\label{DEL}
\tbl{The pentaquark antidecuplet mass spectrum (MeV) in the FS model.}
{\begin{tabular}{@{}cccc@{}}\toprule
${\bf Penta}$ & ${\bf Y,~I,~I_3}$ & ${\bf Present~ results}$ & ${\bf Carlson~ et~ al.}$ \\
               &                           & Ref.~\refcite{FS2}  &   Ref.~\refcite{CARLSON}   \\
\colrule
& & &      \\

${\bf \Theta^+}$ & ${\bf 2,0,0}$  & ${\bf 1540}$ & ${\bf 1540}$   \\

& & &      \\

${\bf N_{\overline {10}}}$      & ${\bf 1,1/2,1/2}$  & ${\bf 1684}$ & ${\bf 1665}$  \\ 

& & &     \\

${\bf \Sigma_{\overline {10}}}$ &  ${\bf 0,1,1}$ & ${\bf 1829}$ & ${\bf 1786}$ \\

& & &      \\

${\bf \Xi^{--}}$ &  ${\bf -1,3/2,-3/2}$ & ${\bf 1962}$ & ${\bf 1906}$ \\

& & &     \\ \botrule
\end{tabular}}
\end{table}
\section{The Wave Function}
It is useful to first look at the $q^4$ subsystem. 
For isospin $ I=0 $ (the $uudd$ system) and spin $S=0$  the lowest
totally antisymmetric state reads 

\begin{equation}  \label{STATE1}
\left|{\psi^{+}}(q^4)\right\rangle =
\left|[{31}]_O
 {\left[{211}\right]}_{C} \left[{{1}^{4}}\right]_{OC}\ ;
[{22}]_{F} [22]_{S} [{4}]_{FS} 
\right\rangle
\end{equation}
which represents the inner product of the orbital (O), color (C), flavor (F)
and spin (S) wave functions  of the $q^4$ subsystem, all written in terms of
partitions $[f]$ associated to various degree of freedom. The $[{4}]_{FS}$ part is
totally symmetric which allows  the maximum possible attraction
in the FS model. It is combined
with the totally antisymmetric $ [{{1}^{4}}]_{OC} $ part, so the total
is an  antisymmetric wave function. The antiquark is then coupled to 
$\left|{\psi^{+}}\right\rangle$ . The symmetry $[31]_O$  requires 
an $s^3 p$ structure i. e. a quark must be  excited to the $p$-shell.
Thus the state (\ref{STATE1}) has positive parity.
One can write such an excited state by using the internal coordinates 

\begin{eqnarray}\label{JACOBI}
\begin{array}{c}\vec{x}\ =\ {\vec{r}}_{1}\ -\ {\vec{r}}_{2}\ , \,
\hspace{5mm} \vec{y}\ =\
{\left({{\vec{r}}_{1}\ +\ {\vec{r}}_{2}\ -\ 2{\vec{r}}_{3}}\right)/\sqrt
{3}}, \nonumber \\
\vec{z}\ =\ {\left({{\vec{r}}_{1}\ +\ {\vec{r}}_{2}\ +\ {\vec{r}}_{3}\ -\
3{\vec{r}}_{4}}\right)/\sqrt {6}}\ , \, \hspace{5mm} \vec{t}\ =\
{\left({{\vec{r}}_{1}\
+\ {\vec{r}}_{2}\ +\ {\vec{r}}_{3}+\ {\vec{r}}_{4}-\
4{\vec{r}}_{5}}\right)/\sqrt {10}}~.
\end{array}
\end{eqnarray}
where 1,2,3 and 4 denote the quarks and 5 the antiquark.
There are 3 independent basis vectors of symmetry $[31]$ or alternatively 
three distinct Young tableaux. These basis vectors can be expressed in terms of 
independent shell model type states $|n \ell m \rangle $ as \cite{FS1}

\begin{eqnarray}\label{psi1}
{\psi }_{1} = \renewcommand{\arraystretch}{0.5}
\begin{array}{c} $\fbox{1}\fbox{2}\fbox{3}$ \\
$\fbox{4}$\hspace{9mm} 
\end{array}
=
\left\langle{\vec{x}\left|{000}\right.}\right
\rangle\left\langle{\vec{y}\left|{000}\right.}\right\rangle\left
\langle{\vec{z}\left|{010}\right.}\right\rangle ~,
\end{eqnarray}

\begin{eqnarray}\label{psi2}
{\psi }_{2} = \renewcommand{\arraystretch}{0.5}
\begin{array}{c} $\fbox{1}\fbox{2}\fbox{4}$ \\
$\fbox{3}$\hspace{9mm} \end{array}
=
\left\langle{\vec{x}\left|{000}\right.}\right
\rangle\left\langle{\vec{y}\left|{010}\right.}\right\rangle\left
\langle{\vec{z}\left|{000}\right.}\right\rangle ~,
\end{eqnarray}

\begin{eqnarray}\label{psi3}
{\psi }_{3} = \renewcommand{\arraystretch}{0.5}
\begin{array}{c} $\fbox{1}\fbox{3}\fbox{4}$ \\
$\fbox{2}$\hspace{9mm} \end{array}
=
\left\langle{\vec{x}\left|{010}\right.}\right
\rangle\left\langle{\vec{y}\left|{000}\right.}\right\rangle\left
\langle{\vec{z}\left|{000}\right.}\right\rangle ~,
\end{eqnarray}
\noindent
This means that the angular excitation $\ell = 1$ can be carried by any
of the relative coordinates $\vec{x}$, $\vec{y}$ or $\vec{z}$, 
with equal probability, as implied by the state (\ref{STATE1}). 
This is entirely different from other pictures promulgated in the literature.
The pentaquark orbital wave function is obtained 
by multiplying each $\psi_i$ by 
$\left\langle{\vec{t}\left|{000}\right.}\right\rangle$ which describes
an $S$-wave state of $\overline {q}$  relative to the $q^4$ subsystem.
Then each orbital wave function becomes a product of four independent
individual wave function, one for each relative coordinate. 
\section{The Light Pentaquark Antidecuplet}

In practice we make a Gaussian Ansatz for each of the individual
wave functions and perform variational calculations. For simplicity we restrict to 
two variational parameters in the 5-body wave 
function, one which we assume to be identical for all three internal
coordinates of  $q^4$
and a different one for the 
relative coordinate of $q^4$ to $\bar q$. The antidecuplet mass spectrum
obtained from such variational calculations is exhibited in Table 1.
Details of the calculations can be found in Ref. ~\refcite{FS2}.
Table 1 compares our results with those of \refcite{CARLSON}.
In the latter, the FS interaction $V_{\chi}$ of (\ref{VCHI}) is reduced
to a form similar to (\ref{CS})
\begin{equation}\label{FS}
 V_{FS} = -  \sum_{i<j}^4 C^{FS}_{ij} \lambda_{i}^{F} \cdot \lambda_{j}^{F}
\vec{\sigma}_i \cdot \vec{\sigma}_j ~.
\end{equation}
Here  $C^{FS}_{ij}$ are radial two-body matrix elements 
specific to the FS model, fitted to reproduce the ground state masses of 
ordinary baryons.
Note that the sum runs over the quarks only. 
In contrast to the present results, in Ref. \refcite{CARLSON} there is
no kinetic term,
no $\eta'$-meson  exchange and no SU(3) breaking   
in the $\eta$-meson exchange due to quark masses. Moreover, the radial two-body
matrix elements do not contain orbital excitation due to the angular 
momentum $\ell = 1$, although the parity is assumed to be positive. 
In both calculations the mass of $\Theta^+$
is fixed to 1540 MeV. One can see that the present calculations lead to larger
splittings between the isomultiplets than those of Ref.~ \refcite{CARLSON}. Then,
in terms of the hypercharge $Y$, they can be parametrized 
by the linear mass formula $M \simeq 1829 - 145~ Y$ while those of 
Ref.~ \refcite{CARLSON}
by $M \simeq 1786 - 122~ Y$.

As mentioned above, the Hamiltonian (\ref{ham})  
breaks the SU(3)$_F$ symmetry.  Thus  mixing of representations appears
naturally. In particular one expects and important mixing between
the ${\overline {10}}_F$ and $8_F$ representations (both
present in the right hand side of 
Eq.~(\ref{REPR})). This implies that octet and antidecuplet states with
identical $Y$, $I$ and $I_3$
should mix. These are $N$ and $\Sigma$ states. The mixing 
leads to physical states, which are either 
``mainly antidecuplet'' or ``mainly octet''. This mixing was discussed in
Ref.~\refcite{FS2} 
where it was found that, besides the free mass term,
a substantial additional contribution to the coupling 
between ${\overline {10}}_F$ and $8_F$ comes 
from the combined effect of the kinetic energy term and the
hyperfine interaction. However the resulting mixing angle was $ 35.34^0$
for N and  - $ 35.48^0$ for $\Sigma$, thus
close to the ideal mixing angle $35.26^0$ for $N$ and -$35.26^0$ 
for $\Sigma$. (We recall that the ideal mixing is due to the
free mass term only.)
Then, for example, the  ``mainly octet'' pentaquark is at 1451 MeV and the 
``mainly antidecuplet'' at 1801 MeV (for details see Ref. ~\refcite{FS2}).

\section {The decay width of $\Theta^+$}
So far there are only schematic studies of the 
strong decay width of $\Theta^+$. On one hand one tries
to attribute the narrowness of the pentaquark resonance
to the smallness of the overlap between a compact $q^4 \bar q$
state and the kinematically allowed final state \cite{CARLSON2,BUCCELLA}.
The size of this overlap results from the algebraic structure of
the wave function (\ref{STATE1}). On the other hand the smallness
of the width is thought to be due to the spatial structure \cite{STECH}
of $\Theta^+$, but there is no dynamical proof of this structure.

\section{The Charmed Antisextet}

The study of the charmed pentaquarks is entirely similar to that
of the light antidecuplet. The essential difference is that the quark-antiquark
interaction can be neglected, due to the heavy mass of the antiquark.

\begin{table}[h]
\label{HEAVY}
\tbl{Masses (MeV)~ of~ the~ positive parity antisextet charmed pentaquarks in various models.}
{\begin{tabular}{@{}ccccccc@{}}\toprule
${\bf Penta}$ & ${\bf I}$ & ${\bf Content}$ & ${\bf FS~ model}$  & ${\bf D-D-\bar c~model}$ 
& ${\bf D-T model}$ & ${\bf Lattice}$ \\
      &         &                           &  Ref.~\refcite{FS1} &  Ref.~\refcite{JW}  & Ref.~\refcite{KL} 
&Ref.~\refcite{CH}     \\
\colrule
& & & & & &   \\

${\bf \Theta^0_c}$ & ${\bf 0}$  & ${\bf u~ u~ d~ d~ \bar c}$ & ${\bf 2902}$ & ${\bf 2710}$ & ${\bf 2985\pm50}$ & ${\bf 2977}$  \\

& & & & & &   \\

${\bf N^0_c}$      & ${\bf 1/2}$  & ${\bf u~ u~d~s~\bar c}$ & ${\bf 3161}$ & & &${\bf 3180}$  \\ 

& & & & & &  \\

${\bf \Xi^0_c}$    &  ${\bf 1}$   & ${\bf u~ u~ s~ s~ \bar c}$ & ${\bf 3403}$ & & &${\bf 3650}$ \\

& & & & & &   \\ \botrule
\end{tabular}}
\end{table}
\noindent
The lowest charmed pentaquarks form an antisextet which is a sub-multiplet 
with charm quantum number $C = - 1$ of the 
${\overline {60}}$ representation of SU(4)$_F$ \cite{WM}. The light antidecuplet and octet
belong to the same representation, but have $C = 0$. The three members of the antisextet
having zero charge are presented in Table 2
together with their quark content. The absolute masses of the antisextet 
members (an exact SU(2) symmetry is assumed), as calculated in the FS model
and extracted from Table II of Ref. \refcite{FS1},  
are compared 
with results recently obtained from lattice calculations, 
Ref.~ \refcite{CH},
where the lowest charmed pentaquarks turn out to have positive parity as well. 
One can see that the FS model and the 
lattice results lead to rather similar predictions.
For completeness we also indicated the only available mass from the 
diquark-diquark-$\bar c$ (D-D-$\bar c)$ model   of Ref. \refcite{JW} and 
that obtained in  Ref. \refcite{KL} in the diquark-triquark (D-T) model.
The mass of $\Theta^0_c$ in the  (D-D-$\bar c)$ model is 
far below the other results.
As mentioned in the introduction, the observation of $\Theta^0_c$
is controversial at present. It is the task of future, perhaps dedicated
experiments, to clarify this situation.

\section{Conclusions}
In regard to the light pentaquarks the conclusion is that the constituent 
quark models can accommodate the pentaquark $\Theta^+$ and its antidecuplet
partners, the lowest antidecuplet states having positive parity and spin 1/2. 
However the absolute mass of $\Theta^+$ cannot be determined. The situation 
is similar to the ordinary baryons where the nucleon ground state is
always fitted to the experimental value.  The mass 
splitting between the antidecuplet isomultiplets has been calculated
in the FS model. 
The mixing between the antidecuplet and octet pentaquarks
was calculated dynamically in the FS model as well and has been found  
close to the ideal mixing. More elaborate five particle
calculations than those based on the variational method used
here are desirable. The calculation  of the decay width should be
performed dynamically.

Manifestly, the existence and properties of the pentaquarks
is a fast moving field.  
It is expected to change at least the usual practice
of baryon spectroscopy. Of particular importance is to understand
the role of the chiral symmetry implemented in the chiral
soliton model where the mass of $\Theta^+$ is so low and its width
is so narrow.
It will be exciting if the pentaquarks will firmly be confirmed
experimentally.

\end{document}